\documentclass[aps,prb,twocolumn,superscriptaddress]{revtex4-2}
\usepackage{graphicx}
\usepackage{amssymb}
\usepackage{amsmath}
\usepackage{multirow}

\begin{document}
\title{Effect of Magnetic Scattering on Superfluid Transition of $^3$He in Nematic Aerogel}
\author{V.\,V.\,Dmitriev}
\email{dmitriev@kapitza.ras.ru}
\affiliation{P.L.~Kapitza Institute for Physical Problems of RAS, 119334 Moscow, Russia}
\author{M.\,S.\,Kutuzov}
\affiliation{Metallurg Engineering Ltd., 11415 Tallinn, Estonia}
\author{A.\,A.\,Soldatov}
\affiliation{P.L.~Kapitza Institute for Physical Problems of RAS, 119334 Moscow, Russia}
\author{A.\,N.\,Yudin}
\affiliation{P.L.~Kapitza Institute for Physical Problems of RAS, 119334 Moscow, Russia}

\date{\today}

\begin{abstract}
We present results of high magnetic field experiments in pure $^3$He (in the absence of $^4$He coverage) in nematic aerogel. In this case the aerogel strands are covered with few atomic layers of solid paramagnetic $^3$He, which enables the spin-exchange mechanism for $^3$He quasiparticles scattering. Our earlier NMR experiments showed that in low fields, instead of the polar phase, the A phase is expected to emerge in nematic aerogel. We use a vibrating wire resonator with the sample of aerogel attached to it and measure temperature dependencies of resonance properties of the resonator at different magnetic fields. A superfluid transition temperature of $^3$He in aerogel, obtained from the experiments, increases nonlinearly in applied magnetic field. And this increase is suppressed compared with that for bulk A$_1$ phase, which we attribute to an influence of the magnetic scattering channel, previously considered theoretically for the case of $^3$He confined in isotropic silica aerogel. However, we observe the essential quantitative mismatch with theoretical expectations.
\end{abstract}

\maketitle

\section{Introduction}
The superfluidity of $^3$He is due to p-wave Cooper pairing with spin and orbital angular momentum equal to 1. This allows the existence of many superfluid phases with different wave functions \cite{VW,Mar}, but only phases with the lowest Ginzburg-Landau free energy are realized. In particular, depending on temperature and pressure, in bulk superfluid $^3$He only two superfluid phases (A and B) exist in zero magnetic field.
Magnetic field changes the energy and A$_1$ phase becomes favorable in a narrow region near the superfluid transition temperature $T_c$. Therefore, instead of the second-order superfluid transition at zero field at $T=T_c$, there are two second-order transitions: the ``upper'' transition to the A$_1$ phase at $T=T_{A1}>T_c$ and the ``lower'' transition to the A$_2$ phase (also called as the A phase in magnetic field) at $T=T_{A2}<T_c$  \cite{Amb,A1,osh,isr84,sag84,koj08}. The A phase is an Equal Spin Pairing (ESP) phase, that is, it contains Cooper pairs with only $\pm$1 spin projections on a specific direction (equal fractions of $\uparrow\uparrow$ and $\downarrow\downarrow$ pairs, where the arrows denote direction of the magnetic moment). The A$_1$ phase contains only $\uparrow\uparrow$ pairs but the A$_2$ phase contains also $\downarrow\downarrow$ pairs, which fraction grows on cooling below $T_{A2}$. In bulk $^3$He, owing to particle-hole asymmetry, the splitting of $T_c$ is proportional to $H$: $T_{A1}=T_c+\eta_{A1} H$ and $T_{A2}=T_c-\eta_{A2} H$, where depending on pressure, $\eta_{A1}=0.6\div4\,\mu$K/kOe and $\eta_{A2}=0.6\div2\,\mu$K/kOe \cite{Amb,isr84,sag84}. Temperature region of existence of the A$_1$ phase is $\Delta T=(\eta_{A1}+\eta_{A2}) H=\eta_A H$.

The similar splitting of the superfluid transition temperature was expected to occur in $^3$He in silica aerogel where the observed A-like phase has the same order parameter as the A phase of bulk $^3$He \cite{aA1,aA2,aA3}. However, experiments with pure $^3$He in silica aerogel show no evidence of the splitting in fields up to 8\,kOe \cite{hal1}, but in very high magnetic fields (greater than 70\,kOe) the splitting linear on $H$ was observed with nearly the same values of $\eta_{A1}, \eta_{A2}$, and $\eta_A$ as in bulk $^3$He \cite{hal2,hal3}. In theoretical works \cite{ss,bh} it was suggested that the observed behavior is due to a magnetic scattering of $^3$He quasiparticles on the aerogel strands. The point is that in pure $^3$He in aerogel the strands of aerogel are covered with $\sim$2 atomic layers of paramagnetic solid $^3$He \cite{Sch,Godf,Coll}. Therefore, during the scattering the spin is not conserved due to a fast exchange between atoms of liquid and solid $^3$He. According to Refs.~\cite{ss,bh}, in presence of the spin exchange the splitting of the superfluid transition temperature in high magnetic fields is suppressed and in paramagnetic model
\begin{equation}
\label{DELT0}
\Delta T=\left(\eta_0-C\frac{\tanh(h)}{h}\right)H,
\end{equation}
where $\eta_0\approx \eta_A T_{ca}/T_c$ is the splitting parameter in absence of the spin exchange, $T_{ca}$ is the superfluid transition temperature of $^3$He in aerogel in zero magnetic field, $h=\gamma \hbar H/(2kT_{ca})$, $\gamma$ is the gyromagnetic ratio, $k$ is the Boltzmann constant, and the spin-exchange parameter $C\sim 1\,\mu$K/kOe depends on the superfluid coherence length, on the mean free path of $^3$He quasiparticles in aerogel, and on impurity scattering parameters.
The ``upper'' transition temperature ($T_{ca1}$) is then given by
\begin{equation}
\label{DELT1}
T_{ca1}=T_{ca}+\left(\eta_{A1}\frac{T_{ca}}{T_c}-C_1\frac{\tanh(h)}{h}\right)H,
\end{equation}
where in the weak coupling limit $C_1=C/2$.

The contribution of the spin-exchange part decreases with the increase of $H$, and for $h\gtrsim 2.5$ the derivatives $dT_{ca1}/dH$
and $d(\Delta T)/dH$ should be nearly equal to $\eta_{A1}T_{ca}/T_c$ and $\eta_0$ respectively as it was observed in experiments described in Ref.~\cite{hal2,hal3}. In lower fields a nonlinear dependence of the splitting on $H$ is expected. For example, if $h\lesssim0.7$ (for $T_{ca}=2$\,mK it corresponds to $H\lesssim20$\,kOe) then from Eq.~\eqref{DELT0} it follows:

\begin{equation}
\label{DELT2}
\Delta T=(\eta_0-C)H +Ch^2H/3=k_1 H+k_2 H^3.
\end{equation}
In experiments described in Ref.~\cite{hal1} no splitting was observed presumably due to $\eta_0$ being nearly equal to $C$, while the second term in Eq.~\eqref{DELT2} at $H<8$\,kOe was still very small.

The effect of magnetic scattering on the splitting should disappear when a small amount of $^4$He is added which replaces the solid layers of $^3$He on the strands. In this case $\Delta T$ should be equal to $\eta_0 H$ at any field. Such experiments in $^3$He in silica aerogel have not been carried out, but they have been done in $^3$He confined in another type of aerogel, that is, in so called nematic aerogel \cite{bet}. In contrast to silica aerogel, which global anisotropy is small, strands of nematic aerogel are aligned on average along the same direction, resulting in a strong anisotropy in the orbital space. In low magnetic fields in presence of the $^4$He coverage such anisotropy makes favorable a new superfluid phase near $T_{ca}$, the polar phase \cite{AI,dmit15}. The polar phase, like the A phase, is the ESP phase and in high magnetic fields the superfluid transition in $^3$He in nematic aerogel is also split \cite{Su1,Su2}: on cooling, the transition occurs into the so-called $\beta$ phase which has the same orbital part of the order parameter as the polar phase, but contains only $\uparrow\uparrow$ pairs. On further cooling, the transition to the distorted $\beta$ phase, which contains also $\downarrow\downarrow$ pairs, is observed. As it was expected, the temperature region of existence of the $\beta$ phase was found to be proportional to $H$ \cite{bet}.

Remarkably, the magnetic scattering essentially influences the superfluid phase diagram of $^3$He in nematic aerogel even at very low magnetic fields: in pure $^3$He (that is in the absence of $^4$He coverage), instead of the transition into the polar phase, the transition to the A phase with substantial suppression of $T_{ca}$ takes place \cite{3he}. Here we describe the experiments in pure $^3$He in nematic aerogel in high magnetic fields. The aim of the experiments was to measure the dependence of the superfluid transition splitting on $H$ and to observe the suppression of the splitting in low fields.
We note that theoretical models \cite{ss,bh} are developed for globally isotropic aerogel and are not applicable directly to $^3$He in nematic aerogel, where the scattering is strongly anisotropic. Nevertheless, the spin exchange mechanism remains and should suppress the splitting in a similar way as it follows from Eqs.~\eqref{DELT1} and \eqref{DELT2} with corrected values of $C$ and $C_1$. In particular, in very high magnetic fields we expected that $dT_{ca1}/dH \approx \eta_{A1}T_{ca}/T_c$.

\section{Samples and methods}
Experiments were performed at a pressure of 15.4\,bar in magnetic fields 2.22--19.36\,kOe using a vibrating wire (VW) resonator with the aerogel sample attached to it. We used the same setup and the same sample of nematic aerogel as in experiments described in Ref.~\cite{bet}.The necessary temperature was obtained by a nuclear demagnetization cryostat and measured using a quartz tuning fork, the resonance linewidth of which in $^3$He depends on temperature. The fork was calibrated as described in Ref.~\cite{bet}.

The sample of mullite nematic aerogel has a form of the rectangular parallelepiped with sizes $\approx2\times3$\,mm transverse to the strands and 2.6\,mm along the strands. Its porosity is 95.2\%, and a characteristic separation between the strands is 60\,nm. The diameter of the strands is $\leq14$\,nm. The sample was glued using a small amount of epoxy resin to 240\,$\mu$m NbTi wire, bent into a shape of an arch with height of 10\,mm and distance between legs of 4\,mm. Strands of the aerogel were oriented along the oscillatory motion (see insert in Fig.~\ref{f1}). The wire is mounted in a cylindrical experimental cell surrounded by a superconducting solenoid, so that the sample is located at the maximum of the magnetic field (with homogeneity of 0.1\% at distances $\pm3$\,mm). A sketch of the cell is shown in Ref.~\cite{bet}. A measurement procedure for the aerogel VW resonator is the same as in the case of a conventional VW resonator \cite{CHH}. The mechanical flapping resonance of the wire is excited by the Lorentz force on an alternating current with amplitude $I_0$ (from 0.2\,mA to 2\,mA depending on $H$ and being set to keep the amplitude of oscillations field independent), passing through the wire. Motions of the wire generate a Faraday voltage which is amplified by a room-temperature step-up transformer {1:30} and measured with a lock-in amplifier by sweeping the frequency. In-phase (dispersion) and quadrature (absorption) signals are joint fitted to Lorentz curves. At $T\sim 1$\,K the resonance frequency and the full width at half-maximum (FWHM) of our VW resonator in vacuum are 621\,Hz and 0.3\,Hz respectively. In liquid $^3$He the maximum velocity of our VW in the used temperature range was always less than  0.2\,mm/s. In a given field additional experiments with a few times smaller excitation currents were also done and showed the same results.

In experiments with the same aerogel VW resonator in the presence of $^4$He coverage it was found that in low magnetic fields at 15.4\,bar $T_{ca}\approx 0.981\,T_c$ \cite{bet}, where $T_c=2.083$\,mK.
This transition occurs into the polar phase as it follows from NMR experiments with similar aerogel sample cut from the same original piece of aerogel \cite{dmit19}. Our previous NMR experiments with pure $^3$He in various samples of nematic aerogel pointed out that in the absence of $^4$He coverage the superfluid transition of $^3$He in the present sample in low magnetic fields should occur to the A phase \cite{3he}. Moreover, $T_{ca}$ is expected to be equal to approximately $0.95\,T_c$ like in pure $^3$He in the sample of nafen-90, because with complete $^4$He coverage the present sample and nafen-90 have basically the same superfluid phase diagrams of $^3$He \cite{3he,dmit19}.

We note that in previous experiments with nematic aerogel VW resonators, performed in the presence of $^4$He coverage, an additional (the second) VW resonance mode was observed, existing only below $T_{ca}$ \cite{bet,vw20}. On cooling from $T=T_{ca}$, the resonant frequency of this additional mode was very rapidly increasing from 0 up to $\sim1.6$\,kHz, and in a narrow temperature range below (but very close to) $T_{ca}$ was reaching the resonance frequency of the main mechanical VW resonance. An interaction of these modes resulted in a peak-like increase of the FWHM of the main mechanical resonance of the VW. Presumably, the second mode is an analog of the second-sound-like mode (slow sound mode) observed in silica aerogel in superfluid helium \cite{McK,gol99} and corresponds to motions in opposite directions of the superfluid component inside the aerogel and the normal component (together with the aerogel strands). In present experiments we focused on measurements of the main resonance, which intensity is significantly greater.

\section{Results and Discussion}
\begin{figure}[t]
\centerline{\includegraphics[width=\columnwidth]{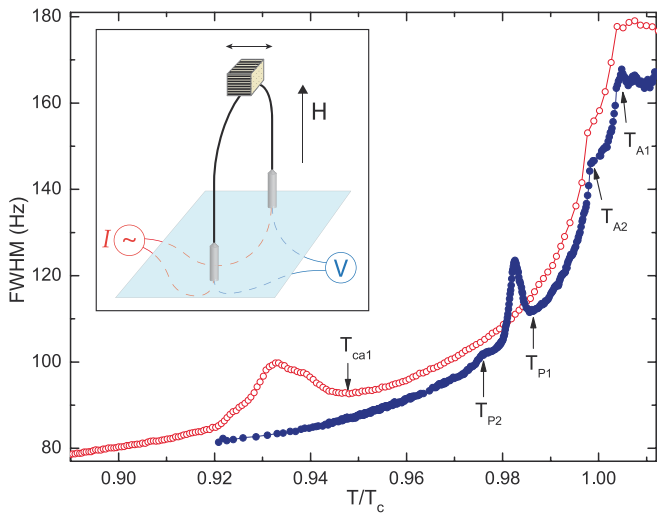}}
\caption{Temperature dependencies of FWHM of the main resonance of the VW resonator measured in the presence of $^4$He coverage (filled circles, $H=4.1$\,kOe) and in pure $^3$He (open circles, $H=4.4$\,kOe). Arrows indicate the features we associate with different superfluid transitions at temperatures $T_{P2}$, $T_{P1}$, $T_{A2}$, $T_{A1}$, and $T_{ca1}$ (see text for details).
The $x$ axis represents the temperature normalized to the superfluid transition temperature in bulk $^3$He $T_c=2.083$\,mK.
(Insert) Signal measurement circuit of a VW immersed in liquid $^3$He in magnetic field $\bf H$. The strands of nematic aerogel are oriented along the oscillations.}
\label{f1}
\end{figure}

In Fig.~\ref{f1} by open circles we show the temperature dependence of the FWHM of the main resonance of the VW obtained in pure $^3$He in magnetic field of 4.4\,kOe. For comparison, we also show by filled circles the similar dependence obtained in experiments described in Ref.~\cite{bet} in the presence of $^4$He coverage at the same pressure and in nearly the same field (4.1\,kOe).

\begin{figure}[t]
\centerline{\includegraphics[width=\columnwidth]{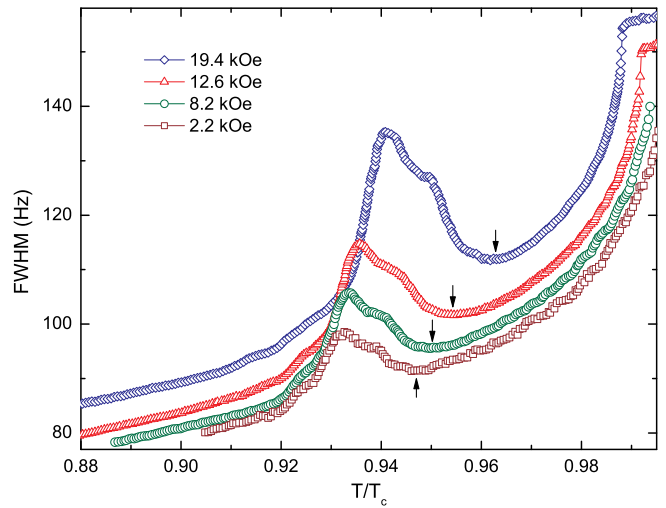}}
\caption{Temperature dependencies of the FWHM of the main resonance of the VW resonator measured in magnetic fields of 2.2\,kOe (squares), 8.2\,kOe (circles), 12.6\,kOe (triangles), and 19.4\,kOe (diamonds) at corresponding excitation currents of 1.9\,mA, 0.5\,mA, 0.33\,mA, and 0.21\,mA. Arrows indicate the features we associate with $T_{ca1}$.}
\label{f2}
\end{figure}

First, let us consider the dependence obtained in the presence of $^4$He coverage.
At $T>T_{A1}$ both bulk $^3$He and $^3$He in aerogel are in the normal state. The superfluid transition to the A$_1$ phase in bulk $^3$He occurs at $T=T_{A1}$. Below $T_{A1}$ the FWHM decreases, and at $T=T_{A2}$ the transition to the A$_2$ phase takes place.
On further cooling, the FWHM decreases more rapidly but below $T=T_{P1}\approx 0.986\,T_c$ it starts to increase that can be due to only the superfluid transition of $^3$He in aerogel. In the given magnetic field it is the transition to the $\beta$ phase. At $T=T_{P2}\approx 0.976\,T_c$ a ``step'' on the FWHM plot is observed, which has been referred to the transition between the $\beta$ phase and the distorted $\beta$ phase existing at $T<T_{P2}$ \cite{bet}. The peak-like change of the FWHM is due to an interaction with the second resonance mode and occurs at $0.98\,T_c<T<0.986\,T_c$. In the case of pure $^3$He we observe bulk A$_1$ and A$_2$ features nearly at the same temperatures $T_{A1}$ and $T_{A2}$, but the superfluid transition of $^3$He inside aerogel occurs at significantly lower temperature $T=T_{ca1}\approx 0.948\,T_c$ than the transition temperature $T_{P1}$ in the presence of $^4$He coverage.
Here we determine the ``upper'' transition temperature $T_{ca1}$ as the temperature corresponding to the local minimum of the dependence of the FWHM on temperature. Such method results in unambiguous determination of $T_{ca1}$ although a systematic error $\sim 0.003\,T_c$ is possible because of a finite temperature width of the superfluid transition of $^3$He in aerogel. We note that in pure $^3$He the peak-like change of the FWHM due to the interaction with the second resonance mode occurs in a rather wide range of temperatures, and presumably the ``lower'' transition temperature $T_{ca2}$ is hard to detect because it is inside this range.

In Fig.~~\ref{f2} we show temperature dependencies of the FWHM of the main VW resonance obtained in different magnetic fields. It is seen that $T_{ca1}$ is increased with the increase of $H$, but we still are not able to detect specific field-dependent features which can be ascribed to transitions at $T=T_{ca2}$.

\begin{figure}[t]
\centerline{\includegraphics[width=1.0\columnwidth]{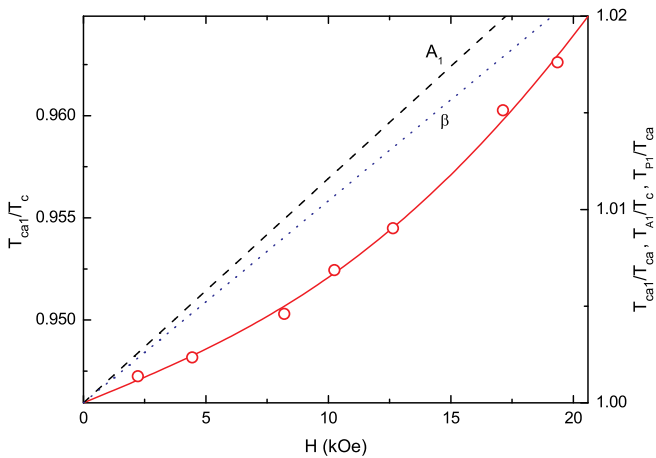}}
\caption{The ``upper'' superfluid transition temperature of $^3$He in nematic aerogel in the absence of $^4$He coverage versus $H$ (circles, left and right axes). The solid line is the best fit of our data by Eq.~\eqref{DELT1} using $\eta_{A1}$ and $C_1$ as fitting parameters. Right axis: the best fit lines for transition temperatures into bulk A$_1$ phase \cite{isr84} (dashed line) and into the $\beta$ phase in $^3$He in the present aerogel sample with $^4$He coverage \cite{bet} (dotted line).}
\label{f3}
\end{figure}

In Fig.~~\ref{f3} we summarize results of our experiments and show the measured dependence of $T_{ca1}$ on $H$. In the same figure we show the best fit lines for transition temperatures into A$_1$ phase ($T_{A1}/T_c$) of bulk $^3$He \cite{isr84} and into the $\beta$ phase ($T_{P1}/T_{ca}$) of $^3$He confined in the present sample of aerogel, but in the presence of $^4$He coverage \cite{bet}.
The solid line is the best fit of our data by Eq.~\eqref{DELT1} using $\eta_{A1}$ and $C_1$ as fitting parameters. Although the fit looks reasonable, it contradicts our expectations. The point is that the value of $\eta_{A1}$ obtained by the fitting is 2.7 times greater than $\eta_{A1}$ in bulk $^3$He. It is also seen that if $H\gtrsim$15\,kOe then the derivative $dT_{ca1}/dH$ exceeds the value of the derivative $dT_{A1}/dH$ while it should only reach this value in magnetic fields of $H\gtrsim 70$\,kOe ($h\gtrsim 2.5$).

Further theoretical investigations, as well as experiments in very high magnetic fields, are necessary to understand the observed discrepancy with theoretical expectations. Here we can suggest only one naive explanation of this discrepancy.
In the absence of the magnetic scattering (i.e., in the presence of $^4$He coverage) the superfluid transition in $^3$He in nematic aerogel in low magnetic fields occurs into the polar phase with a rather small suppression of $T_{ca}$ with respect to $T_c$ \cite{dmit15}. In high magnetic fields the transition occurs into the $\beta$ phase \cite{bet}. Both polar and $\beta$ phases become more favorable, rather than A and A$_1$ phases, due to a strong anisotropy of $^3$He quasiparticles scattering within nematic aerogel. However, the magnetic scattering, which is enabled in pure $^3$He, changes the Ginzburg-Landau free energy and makes favorable A and A$_1$ phases with a substantial suppression of $T_{ca}$. For $^3$He in nematic aerogel, the free energy should contain, firstly, field-dependent contribution including interference term between the spin-independent and spin-exchange parts of quasiparticle scattering from spin-polarized $^3$He on the aerogel strands (like it was considered in the case of isotropic silica aerogel in Refs.~\cite{ss,bh}), secondly, terms accounting for strongly anisotropic impurities. So, in the presence of a magnetic field not only a superfluid transition temperature, but also an order parameter itself is a subject to change if it leads to a minimum of the free energy. Therefore, it is possible that in pure $^3$He in high magnetic fields the polar-distorted A$_1$ phase (or even $\beta$ phase) may become more favorable than the A$_1$ phase. The order parameter of the polar-distorted A$_1$ phase is: $A_{\mu j}=\Delta (d_\mu+ie_\mu) (am_j+ibn_j)$, where $\Delta$ is the gap parameter, $\bf d$ and $\bf e$ are mutually orthogonal unit vectors in spin space, $\bf m$ and $\bf n$ are mutually orthogonal unit vectors in orbital space, and $a^2+b^2=1$. In pure A$_1$ phase $a=b$, in pure $\beta$ phase $a=1, b=0$, and in the polar-distorted A$_1$ phase $a>b$. Thus we make an assumption that in our experiments the magnetic field indirectly changes the orbital part of the order parameter of the observed superfluid phase of $^3$He in nematic aerogel: in very low fields the superfluid transition occurs to the A$_1$ phase with $a=b$, but in higher magnetic fields the A$_1$ phase acquires a polar distortion (the ratio $a/b$ is increased) that should be accompanied by a corresponding increase in $T_{ca}$ \cite{JLT,Fom}. As a result, we obtain this additional increase of $T_{ca1}$ from Eq.~\eqref{DELT1}.

\section{Conclusions}
Using the VW techniques in magnetic fields up to 20\,kOe, we have performed experiments in $^3$He confined in nematic aerogel in the absence of $^4$He coverage and have measured the field dependence of the superfluid transition temperature of $^3$He in aerogel. It was found that this dependence is nonlinear and that the increase of the transition temperature with the increase of $H$ is suppressed in comparison with the dependence for bulk A$_1$ phase. We ascribe this suppression to an influence of the magnetic scattering on the splitting of the superfluid transition temperature as it was proposed in theoretical works \cite{ss,bh}. However, we obtain the essential quantitative mismatch with theoretical expectations. This mismatch may be explained in assumption that the superfluid order parameter depends on $H$ due to a possible field dependence of the influence of the magnetic scattering on the Ginzburg-Landau free energy.

\begin{acknowledgments}
The aerogel sample was made by M.S.K. The experiments was carried out by V.V.D., A.A.S., and A.N.Y. and supported by grant of the Russian Science Foundation (project \#\,18-12-00384). We are grateful to I.A.~Fomin and E.V.~Surovtsev for useful discussions and comments.
\end{acknowledgments}

\end{document}